# Deep-learning continuous gravitational waves

Christoph Dreissigacker,[1,2,*] Rahul Sharma,[3,1,2] Chris Messenger,[4] Ruining Zhao,[5,6,7] and Reinhard Prix[1,2]

[1]*Max Planck Institute for Gravitational Physics (Albert-Einstein-Institute), D-30167 Hannover, Germany*
[2]*Leibniz Universität Hannover, D-30167 Hannover, Germany*
[3]*Birla Institute of Technology and Science, Pilani, Rajasthan 333031, India*
[4]*SUPA, School of Physics and Astronomy, University of Glasgow, Glasgow G12 8QQ, United Kingdom*
[5]*Key Laboratory of Optical Astronomy, National Astronomical Observatories,
Chinese Academy of Sciences, Beijing 100101, China*
[6]*University of Chinese Academy of Sciences, Beijing 100049, China*
[7]*Department of Astronomy, Beijing Normal University, Beijing 100875, China*

We present a first proof-of-principle study for using deep neural networks (DNNs) as a novel search method for continuous gravitational waves (CWs) from unknown spinning neutron stars. The sensitivity of current wide-parameter-space CW searches is limited by the available computing power, which makes neural networks an interesting alternative to investigate, as they are extremely fast once trained and have recently been shown to rival the sensitivity of matched filtering for black-hole merger signals [D. George and E. A. Huerta, Phys. Rev. D **97**, 044039 (2018); H. Gabbard, M. Williams, F. Hayes, and C. Messenger, Phys. Rev. Lett. **120**, 141103 (2018)]. We train a convolutional neural network with residual (shortcut) connections and compare its detection power to that of a fully coherent matched-filtering search using the WEAVE pipeline [K. Wette, S. Walsh, R. Prix, and M. A. Papa, Phys. Rev. D **97**, 123016 (2018)]. As test benchmarks we consider two types of all-sky searches over the frequency range from 20 to 1000 Hz: an "easy" search using $T = 10^5$ s of data, and a "harder" search using $T = 10^6$ s. The detection probability $p_{\text{det}}$ is measured on a signal population for which matched filtering achieves $p_{\text{det}} = 90\%$ in Gaussian noise. In the easiest test case ($T = 10^5$ s at 20 Hz) the DNN achieves $p_{\text{det}} \sim 88\%$, corresponding to a loss in sensitivity depth of $\sim 5\%$ versus coherent matched filtering. However, at higher frequencies and for longer observation times the DNN detection power decreases, until $p_{\text{det}} \sim 13\%$ and a loss of $\sim 66\%$ in sensitivity depth in the hardest case ($T = 10^6$ s at 1000 Hz). We study the DNN generalization ability by testing on signals of different frequencies, spindowns and signal strengths than they were trained on. We observe excellent generalization: only five networks, each trained at a different frequency, would be able to cover the whole frequency range of the search.

DOI: 10.1103/PhysRevD.100.044009

## I. INTRODUCTION

Gravitational waves from binary mergers are now being observed routinely [1–4] by the Advanced LIGO [5] and Virgo [6] detectors. In contrast, the much weaker and longer-lasting (days–months) narrow-band *continuous gravitational waves* (CWs) from spinning deformed neutron stars are yet to be detected, despite a multitude of searches over the past decade (see Refs. [7–9] for reviews) and continuing improvements in search methods (see e.g., Ref. [10] for a recent overview).

A key limitation of current search methods for CWs with unknown parameters is the "exploding computing cost problem": give that a putative signal would be very weak, one needs to integrate as much data as possible in order to increase the signal-to-noise ratio (SNR). However, for a fully coherent matched-filtering search (which is close to statistically optimal [11]), the corresponding computing cost grows as a high power $\sim T^n$ of the data time span $T$, with $n \gtrsim 5$. This typically limits the longest coherent duration to days–weeks before the computing cost would become infeasible.

Therefore the class of semicoherent methods has been developed, producing computationally cheaper searches. They allow the analysis of more data, typically resulting in

---

better sensitivity than a corresponding coherent search at fixed computing cost (see e.g., Refs. [12,13]). Such methods combined with massive amounts of computing power, either via local computer clusters or via the distributed public computing platform Einstein@Home [14], currently yield the best state-of-the-art sensitivity to CW signals (see e.g., Refs. [15–17] for recent examples).

In this work we investigate *deep neural networks* (DNNs) [18–20] as a novel search method for CWs. The field of DNNs, also referred to as *deep learning*, has emerged as one of the most successful machine-learning paradigms in the last decade, dominating wide-ranging fields [20] such as image recognition, speech recognition and language translation, as well as certain board [21] and video games [22,23].

More recently DNNs have started to draw attention in the field of gravitational-wave searches (i) as a classifier for non-Gaussian detector transients (*glitches*) [24–27], (ii) as a search method for unmodeled burst signals [28,29] in time-frequency images produced by coherent WaveBurst [30], and (iii) as a direct detection method for black-hole merger signals in gravitational-wave strain data [31–36].

This last approach (iii) is of particular interest to us, as Refs. [31,32] illustrated for the first time that DNNs can achieve a detection power comparable to that of (near-optimal) matched filtering, at a fraction of the search time. This is relevant for CW searches: while semicoherent methods for wide-parameter-space searches are the most sensitive approach currently known, they are by design less sensitive than the statistical optimum achievable according to the Neyman-Pearson-Searle lemma [37].

With DNNs the computationally expensive step is shifted to the *preparation* stage of the search—the architecture tuning and "learning" of optimal network weights (i.e., the *training*)—while the execution time on given input vectors is very short (typically fractions of a second). The determination of the noise distribution (for estimation of the false-alarm level $p_{\text{fa}}$) and measurement of upper limits require many repeated searches over different input data sets, with and without injected signals. The relative search speed advantage of DNNs compared to traditional search methods therefore accumulates dramatically over these operations allowing very fast and flexible search characterizations.

The plan of this paper is as follows. In Sec. II we define and characterize our test benchmarks. In Sec. III we describe our deep-learning approach to searching for continuous gravitational waves, and explain the network architecture and how it was trained. In Sec. IV we characterize the performance our DNN achieves on the test benchmarks in comparison to the matched-filtering performance and how it generalizes beyond the benchmarks' search parameters. And finally we discuss these results in Sec. V.

## II. COMPARISON TEST BENCHMARKS

### A. Benchmark definitions

In order to characterize the detection power of the DNN that we introduce in the next section, we define two benchmark search setups and measure the corresponding sensitivity achieved on them with a classical (near-optimal) matched-filter search method described in Sec. II B.

We compare the sensitivity in the Neyman-Pearson sense, also known as the receiver-operator characteristic (ROC), using the "upper limit" conventions used in most CW searches (cf. Ref. [10]): measure the detection probability $p_{\text{det}}$ at a chosen false-alarm level $p_{\text{fa}}$ for a signal population of fixed amplitude $h_0$, with all other signal parameters (i.e., polarization, sky position, frequency and spindown) drawn randomly from their priors. In order to characterize the signal strength in noise, we use the *sensitivity depth* $\mathcal{D}$ [10,38], defined as

$$\mathcal{D} \equiv \frac{\sqrt{S_n}}{h_0}, \quad (1)$$

where $S_n$ is the power spectral density of the (Gaussian) noise at the signal frequency, and $h_0$ is the signal amplitude. In particular we are interested in the sensitivity depth $\mathcal{D}^{90\%}$ that corresponds to the signal amplitude $h_0^{90\%}$ at which the search yields a detection probability of $p_{\text{det}} = 90\%$ at a fixed false-alarm level, which here is chosen as $p_{\text{fa}} = 1\%$ per 50 mHz frequency band.

We choose to use stationary white Gaussian noise, which allows us to efficiently generate training data, and it simplifies comparing against the sensitivity of idealized matched filtering.

We consider two *all-sky* searches (parameters summarized in Table I) over a range of frequency $f$ and first-order spindown $\dot{f}$, one using $T = 10^5$ s $\sim 1.2$ days, and one using $T = 10^6$ s $\sim 12$ days of data assuming a single detector (chosen as LIGO Hanford). These two searches could realistically be performed with coherent matched filtering. The required computing cost for the search and its characterization (upper limits, false-alarm level) however would still require a large cluster of, say, $\mathcal{O}(1000)$ cores for over a month or so (see Table II). Therefore actually performing these two full searches only for the purpose of characterizing the matched-filtering sensitivity would be infeasible. Instead we characterize the matched-filter search on only five narrow frequency bands of width $\Delta f = 50$ mHz starting at frequencies $f_0 = 20, 100, 200, 500$ and 1000 Hz, yielding a total of ten representative test cases.

TABLE I. Definition of the two benchmark searches.

| | |
|---|---|
| Data span | $T = 10^5$ s$/T = 10^6$ s |
| Detectors | LIGO Hanford |
| Noise | Stationary, white, Gaussian |
| Sky-region | All-sky |
| Frequency band | $f \in [20, 1000]$ Hz |
| Spindown range | $\dot{f} \in [-10^{-10}, 0]$ Hz/s |

TABLE II. WEAVE parameters and characteristics for the two searches.

| Name | $T = 10^5$ s | $T = 10^6$ s |
|---|---|---|
| Mismatch parameter $m$ | 0.1 | 0.2 |
| Average SNR loss $\langle\mu\rangle$ | 5% | 11% |
| Number of templates $\mathcal{N}$ | $4 \times 10^{11}$ | $3 \times 10^{14}$ |
| Search time [single CPU core] | $6.7 \times 10^6$ s | $3.9 \times 10^9$ s |

### B. WEAVE matched-filtering sensitivity

For the matched-filter search we use the recently developed WEAVE code [39], which implements a state-of-the-art CW search algorithm [40] based on summing coherent $\mathcal{F}$ statistics [41,42] over semicoherent segments on optimal lattice-based template banks [43,44]. This code can also perform fully coherent (i.e., single-segment) $\mathcal{F}$-statistic searches, which we use for the present proof-of-principle study. The benchmark search definitions in Table I are chosen in such a way that a fully coherent search is still computationally feasible. This yields a simpler and cleaner comparison than using a semicoherent search setup, as we can easily design near-optimal search setups (by using relatively fine template banks) without the extra complication of requiring costly sensitivity optimization at fixed computing cost [13,40,45].

The WEAVE template banks are characterized by a maximal-mismatch parameter $m$, which controls how fine the templates are spaced in parameter space. These are chosen as $m = 0.1$ and $m = 0.2$ for the two searches with $T = 10^5$ and $T = 10^6$ s, respectively. The reason for choosing the larger mismatch value (i.e., coarser template bank) in the $T = 10^6$ s case is to keep the computing cost of the corresponding test cases still practically manageable, as the coherent cost scales with the mismatch parameter as $\propto m^2$ for a four-dimensional template bank [see e.g., Eq. (24) in [43]].

By repeated injections of signals in the data and recovery of the loudest $\mathcal{F}$-statistic candidate in the template bank, one can measure the relative SNR loss $\mu$ compared to a perfectly matched template. The resulting measured average mismatch $\langle\mu\rangle$ quantifies in some sense how close to "optimal" the matched-filter sensitivity will be (compared to an infinite computing cost search with $m = 0$), and is found as $\langle\mu\rangle \sim 5\%$ and $\langle\mu\rangle \sim 11\%$, respectively for the two searches.

Using the template-counting and timing models [39,46,47] for WEAVE and the resampling $\mathcal{F}$ statistic, we can estimate the total number of templates and the corresponding total runtime for these two benchmark searches as ~78 and ~45000 days on a single CPU core, respectively. Table II provides a summary of the WEAVE search parameters and characteristics.

In order to estimate the sensitivity for the ten test cases defined in the previous section (i.e., five frequency slices of $\Delta f = 50$ mHz for each search of $T = 10^5$ and $T = 10^6$ s, respectively), we first determine the corresponding detection thresholds $\mathcal{F}_{\text{th}}$ on the $\mathcal{F}$ statistic corresponding to a false-alarm level of $p_{\text{fa}} = 1\%$ for each case. This is done by repeatedly ($10^5$ times for $T = 10^5$ s, and $\sim 10^4$ times for $T = 10^6$ s, respectively) performing each search over Gaussian noise and thereby recording the distribution of the loudest candidate, which yields the relationship between the threshold and false-alarm level. The corresponding detection probability $p_{\text{det}}$ for any given signal population of fixed $\mathcal{D}$ is obtained by injecting signals into Gaussian noise data and measuring how many times the loudest candidate exceeds the detection threshold. By varying the injected $\mathcal{D}$ we can eventually find $\mathcal{D}^{90\%}$ for the desired $p_{\text{det}} = 90\%$ (see e.g., Ref. [10] for more details and discussion of this standard "upper limit" procedure). By a final injection + recovery Monte Carlo we can verify that the achieved WEAVE detection probability for the quoted thresholds and signal strengths $\mathcal{D}^{90\%}$ in Table III is $p_{\text{det}} \sim 90$–91%, which is sufficiently accurate for our present purposes.

The sky template resolution grows as $\propto f^2$ as a function of frequency $f$, resulting in a corresponding increase in the number of templates at higher frequency. This increases the number of "trials" in noise at the higher-frequency slices, which results in a corresponding increased false-alarm threshold (chosen in order to keep the false-alarm level at $p_{\text{fa}} = 1\%$) as well as an increased computing cost, shown in Table III.

TABLE III. WEAVE characteristics for the ten test cases, each covering a frequency "slice" of $\Delta f = 50$ mHz, starting at $f_0$, of the full searches defined in Table I. The detection thresholds $\mathcal{F}_{\text{th}}$ correspond to a false-alarm level of $p_{\text{fa}} = 1\%$ over the band $\Delta f$. $\text{CPU}_{\Delta f}$ denotes the search time in seconds for the respective $\Delta f$ band on a single CPU core.

| $f_0$ | 20 Hz | 100 Hz | 200 Hz | 500 Hz | 1000 Hz |
|---|---|---|---|---|---|
| $T = 10^5$ s | | | | | |
| $\mathcal{N}_{\Delta f}$ | $5 \times 10^5$ | $1 \times 10^7$ | $5 \times 10^7$ | $3 \times 10^8$ | $1 \times 10^9$ |
| $\text{CPU}_{\Delta f}$ [s] | 0.1 | 4.9 | 19 | $2.3 \times 10^2$ | $1.7 \times 10^3$ |
| $\mathcal{F}_{\text{th}}(p_{\text{fa}})$ | 20.6 | 23.6 | 25.1 | 27.0 | 28.6 |
| $T = 10^6$ s | | | | | |
| $\mathcal{N}_{\Delta f}$ | $3 \times 10^8$ | $8 \times 10^9$ | $3 \times 10^{10}$ | $2 \times 10^{11}$ | $8 \times 10^{11}$ |
| $\text{CPU}_{\Delta f}$ [s] | 45 | $3 \times 10^3$ | $1.4 \times 10^4$ | $1.6 \times 10^5$ | $6.9 \times 10^5$ |
| $\mathcal{F}_{\text{th}}(p_{\text{fa}})$ | 27.5 | 31.1 | 32.5 | 34.2 | 36.2 |

## III. DEEP-LEARNING CWs

Our general approach is similar to that of Refs. [31,32] in that we directly use the detector strain data as our network input, and train a simple classifier with two output neurons for the classes "noise" and "signal (in noise)." However, given that CW signals are long in duration and narrow in frequency, instead of using the time-series input it makes more sense in our case to use the frequency-domain representation of that data. We therefore provide the real and imaginary parts of the fast Fourier transform (FFT) of the data as a two-dimensional input vector over frequency bins, using the native FFT resolution of $1/T$. We chose the network input size to be sufficiently large to contain the widest signal (signals get stretched in the frequency domain by the spindown $\dot{f}$ and Doppler shifts) *twice*, so that we can slide the network along the frequency axis in steps of half the network input width, guaranteeing that one input window will always contain the full signal.

### A. Network architecture

We started experimenting with DNN architectures similar to those described in Refs. [31,32], but eventually by trial and error converged on a ResNet architecture [48], which showed better performance for our problem cases.

We have chosen slightly different networks for the two searches ($T = 10^5$ and $T = 10^6$ s) of Table I, as these correspond to signals with rather different widths in the frequency domain: the network in the $T = 10^5$ s cases contains six instances of a residual block, while in the $T = 10^6$ s cases the network uses 12.

The network layers can be separated into three parts: the stem block, a block of multiple residual blocks, and an end block; see Fig. 1. The stem block consists of a standard convolutional layer, while each of the residual blocks is built according to Ref. [48]. The end block contains a dense softmax layer with two final output neurons, corresponding to the estimated probability $p_{\text{signal}}$ that the input contains a signal, and $p_{\text{noise}} = 1 - p_{\text{signal}}$ for a pure noise sample. The DNNs are implemented in the KERAS framework [49] on top of a TENSORFLOW [50] backend.

### B. DNN training and validation

Training the network is performed on a synthesized data set of input vectors, where half contain pure Gaussian noise, and half contain a signal added to the noise. One full pass through this training set is commonly referred to as a training *epoch*. Using a precomputed set of 10 000 signals, each signal is added to 24 dynamically generated noise realizations, which are also used as pure-noise inputs.

The number of signals in the training set is chosen by observing that in all our cases we find that going beyond 10 000 signals yields only negligible further improvements in detection probability, as shown in Fig. 2 for the example of $T = 10^5$ s, $f_0 = 1000$ Hz.

The signals are scaled to a fixed depth $\mathcal{D}^{90\%}_{\text{training}}$ for each test case and randomly shifted in frequency within the network input window. These training depths were estimated semianalytically using the method of Refs. [10,47], and differ slightly from the final measured values $\mathcal{D}^{90\%}$ of Table IV, which were not available at the time of training. When testing the network on signals of different depths, the detection probability behaves as expected; see Sec. IV D. Furthermore, we found that using a different choice of training depth did not significantly affect training success.

Every few epochs of training, we perform a *validation* step, where the detection probability of the network is measured on an independent data set. This validation set contains another 20 000 input vectors, with half containing signals in noise (of fixed depth $\mathcal{D}^{90\%}$), and half containing noise only.

In order to compute the network's detection probability $p^{\text{DNN}}_{\text{det}}$, we treat the output neuron $p_{\text{signal}}$ as a statistic, and follow the usual "upper limit" procedure described in

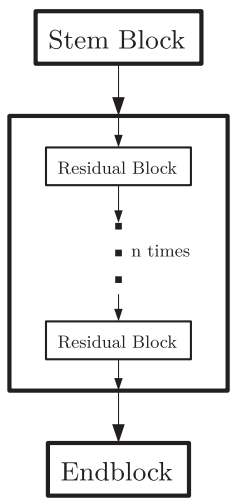

FIG. 1. Illustration of the general network architecture used in this study.

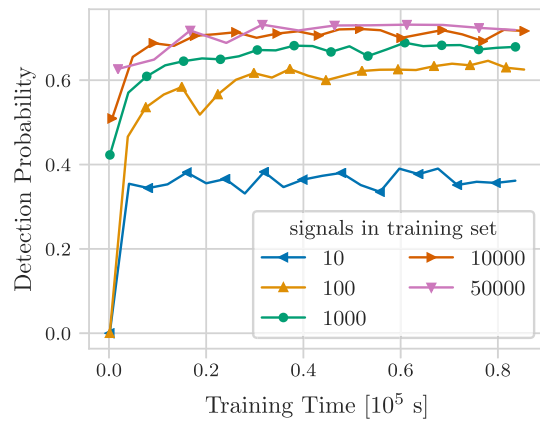

FIG. 2. Validation detection probability for $T = 10^5$ s, $f_0 = 1000$ Hz for training with training sets containing 10, 100, 1000, 10 000 and 50 000 signals.

TABLE IV. Measured WEAVE "upper limit" sensitivity $\mathcal{D}^{90\%}$ at a false-alarm level of $p_{\mathrm{fa}} = 1\%$.

| $\mathcal{D}^{90\%}$ [Hz$^{-1/2}$] | $f_0 = 20$ Hz | 100 Hz | 200 Hz | 500 Hz | 1000 Hz |
|---|---|---|---|---|---|
| $T = 10^5$ s | 11.4 | 10.8 | 10.4 | 9.9 | 9.7 |
| $T = 10^6$ s | 29.3 | 28.2 | 27.6 | 26.8 | 26.0 |

Sec. II B: we repeatedly run the network on Gaussian noise inputs in order to determine the $p_{\mathrm{fa}} = 1\%$ detection threshold. We then run the network on the signal set and measure for what fraction of signals the statistic exceeds that threshold.

The evolution of the detection probability as a function of training epoch (or similarly, as a function of training time) is presented in Fig. 3, illustrating the progress of learning. In order to test the variability and dependence of the learning success on the random initialization of the network, we train a "cloud" of ∼100 differently initialized network instances. We use the network at its point of best validation performance from each test case for the further test results presented in the next sections.

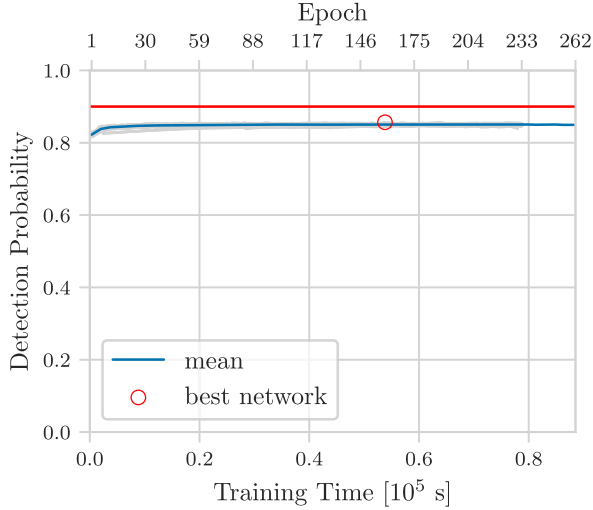
(a) $T = 10^5$ s, $f_0 = 20$ Hz

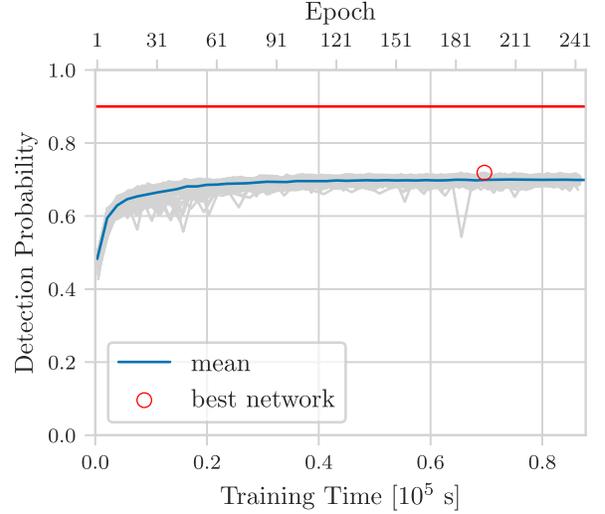
(b) $T = 10^5$ s, $f_0 = 1000$ Hz

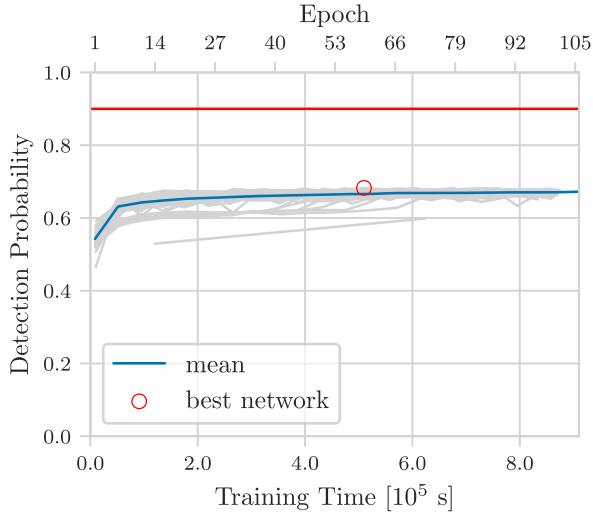
(c) $T = 10^6$ s, $f_0 = 20$ Hz

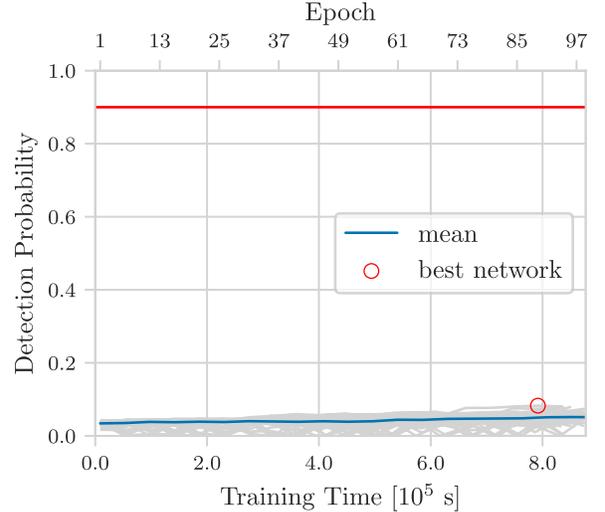
(d) $T = 10^6$ s, $f_0 = 1000$ Hz

FIG. 3. Validation detection probability $p_{\mathrm{det}}^{\mathrm{DNN}}$ of the DNN versus training time (or mean trained epoch) for 100 different network instances trained for each of four test cases: (a) $T = 10^5$ s, $f_0 = 20$ Hz, $p_{\mathrm{det}}^{\mathrm{best}} = 85.7\%$; (b) $T = 10^5$ s, $f_0 = 1000$ Hz, $p_{\mathrm{det}}^{\mathrm{best}} = 71.9\%$; (c) $T = 10^6$ s, $f_0 = 20$ Hz, $p_{\mathrm{det}}^{\mathrm{best}} = 68.3\%$; and (d) $T = 10^6$ s, $f_0 = 1000$ Hz, $p_{\mathrm{det}}^{\mathrm{best}} = 8.3\%$, all trained on an Nvidia GTX 750. The solid horizontal line denotes the matched-filtering detection performance of $p_{\mathrm{det}} = 90\%$. The red circles indicate the networks with the best detection probability $p_{\mathrm{det}}^{\mathrm{best}}$. The error bars on each of the 100 curves are smaller than the widths of the lines.

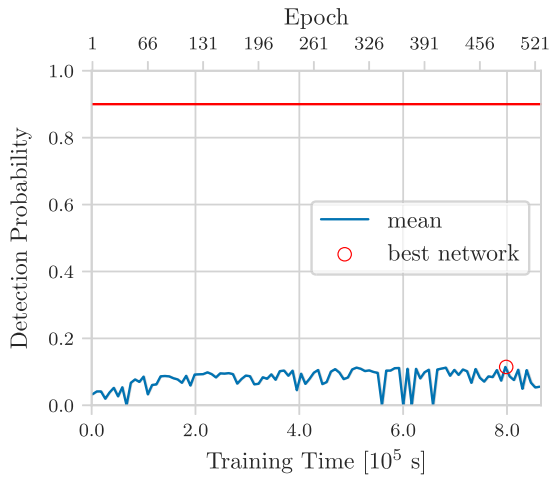

FIG. 4. Validation detection probability $p_{\text{det}}^{\text{DNN}}$ of the DNN versus training time for a single network trained on an Nvidia TITAN V for the case $T = 10^6$ s, $f_0 = 1000$ Hz. The best network, indicated by the red circle, achieves a detection probability of $p_{\text{det}}^{\text{best}} = 11.5\%$. The error bars are smaller than the widths of the lines.

Most of the training was performed on Nvidia GTX 750 GPUs. We see in Fig. 3 that for most cases the improvements in detection probability seem to level off after the training time (about one day in the $T = 10^5$ s cases, and about 10 days in the $T = 10^6$ s cases). However, in the case of $T = 10^6$ s, $f_0 = 1000$ Hz seen in Fig. 3(d) [and also for $T = 10^6$ s, $f_0 = 500$ Hz (not shown)], there still seems to be a slowly increasing trend in detection probability at the end of training time. Therefore we trained a single network instance for these two cases again on a more powerful Nvidia TITAN V GPU for many more epochs, until the validation detection probability seemed to level off, which is shown in Fig. 4.

Overall we observe a dramatic increase in the "difficulty" the DNN has in learning the different test cases along the direction of increasing data span $T$ and frequency $f$, also seen clearly in Table V. In the easiest case of $T = 10^5$ s, $f_0 = 20$ Hz the DNN achieves a detection probability of $p_{\text{det}}^{\text{DNN}} \sim 88\%$, nearly rivaling matched-filtering performance, while in the hardest case of $T = 10^6$ s, $f_0 = 1000$ Hz it only manages $p_{\text{det}} \sim 13\%$ (also see Table V). This may not be very surprising, given that the cases become increasingly more computationally intensive (more templates) along the same axis for matched filtering, as seen in Table III. In the frequency-domain input vectors of the DNN, this would manifest by the signals being more widely spread out due to increased frequency drift $\dot{f}T$ and Doppler stretching.

TABLE V. Detection probabilities in % of the best networks for each case at a false-alarm level $p_{\text{fa}} = 1\%$ and 90% matched-filtering depth.

| $p_{\text{det}}^{90\%}$ | $f_0 = 20$ Hz | 100 Hz | 200 Hz | 500 Hz | 1000 Hz |
|---|---|---|---|---|---|
| $T = 10^5$ s | $87.6^{+0.7}_{-0.6}$ | $85.4^{+0.7}_{-0.7}$ | $84.1^{+0.7}_{-0.7}$ | $80.2^{+0.8}_{-0.8}$ | $73.0^{+0.9}_{-0.9}$ |
| $T = 10^6$ s | $68.8^{+0.9}_{-0.9}$ | $50.0^{+1.0}_{-1.0}$ | $38.7^{+0.9}_{-1.0}$ | $25.4^{+0.8}_{-0.9}$ | $13.1^{+0.6}_{-0.7}$ |

## IV. TESTING DNN PERFORMANCE

After the training and validation steps, we perform a series of tests on the best DNN found for each test case (i.e., with the highest $p_{\text{det}}^{\text{DNN}}$ over all validation steps), in order to more fully characterize its performance as a CW detection method. In these tests we simulate the signals and noise directly for any given depth using the standard CW LALSUITE [51] machinery, in order to independently verify the network performance. Hence we are not using a traditionally fixed *testing set* but generate it on demand.

### A. Verifying detection probabilities

As a sanity check we use an independent test pipeline to confirm the detection probabilities $p_{\text{det}}^{\text{DNN}}$ for the ten cases. These results, given in Table V, are seen to be between 0.5–2 percentage points higher than the corresponding validation $p_{\text{det}}^{\text{best}}$ originally observed in Figs. 3 and 4. This can be understood as follows: in order to speed up training, in the validation step we do not slide the network window over the search frequency $\Delta f$ in the signal case, but instead use only one network window fully containing the signal. In the test case, on the other hand, we perform a more accurate simulation of a real search by sliding (see Sec. III), which can only increase the detection probability.

A second interesting question is how the detection probability depends on the false-alarm level $p_{\text{fa}}$ (commonly referred to as ROC curve) for a fixed signal population. This is shown in Fig. 5 in comparison to the matched-filter ROC.

### B. Generalization in frequency $f$

If we want to perform a search over the whole frequency range (e.g., as defined in Table I) using DNNs, we would need to determine how many different networks we have to train in order to cover this range with a reasonable overall sensitivity. Alternatively we can also train a single DNN with signals drawn from the full frequency range of the search and compare its performance.

The results of these tests are shown in Fig. 6, which show how the five DNNs, trained at their respective frequencies $f_0$, perform over the full frequency range of the search. In addition we show the performance of another network that has been trained directly over the full frequency range.

We see that the "specific" networks trained only on a narrow frequency range still perform reasonably well over a fairly broad range of frequencies, and especially that networks trained at higher frequencies generalize well to lower frequencies. This result shows that a small number of networks $\mathcal{O}(5)$ would be able to cover the whole frequency range at a similar detection performance that was obtained on the individual training frequencies. Furthermore, for the $T = 10^5$ s search, it seems quite feasible to train a single network over the full frequency range directly, achieving similar (albeit lower) performance to the "specialized" networks trained on narrow frequency bands. On the contrary for the $T = 10^6$ s search the detection probability of the "full-range" network drops up to 20 percentage points against the "specialized" networks.

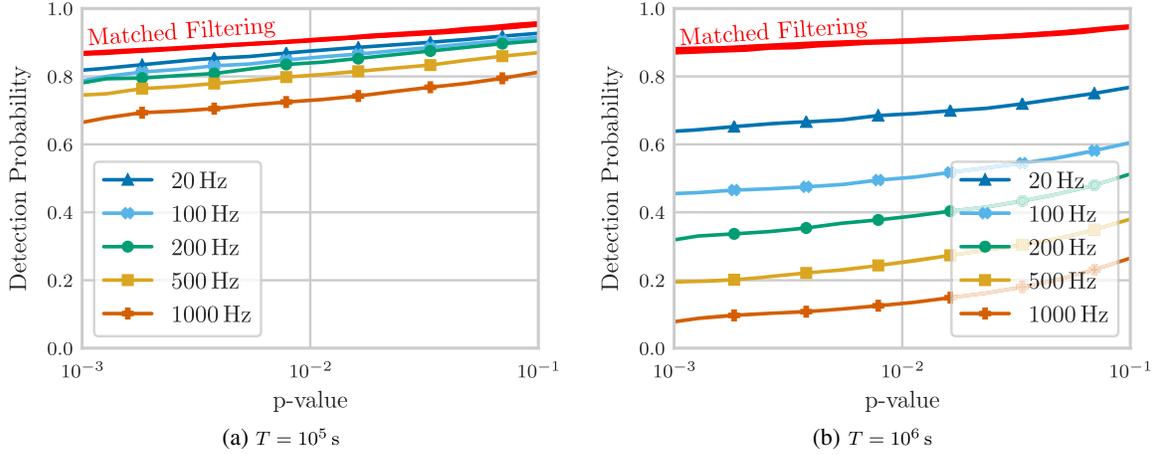

FIG. 5. ROC curve: Detection probability $p_{\text{det}}$ versus $p_{\text{fa}}$ for the $10^5$ s search (left) and the $10^6$ s search (right). The solid red lines indicate the measured ROC curves for matched filtering.

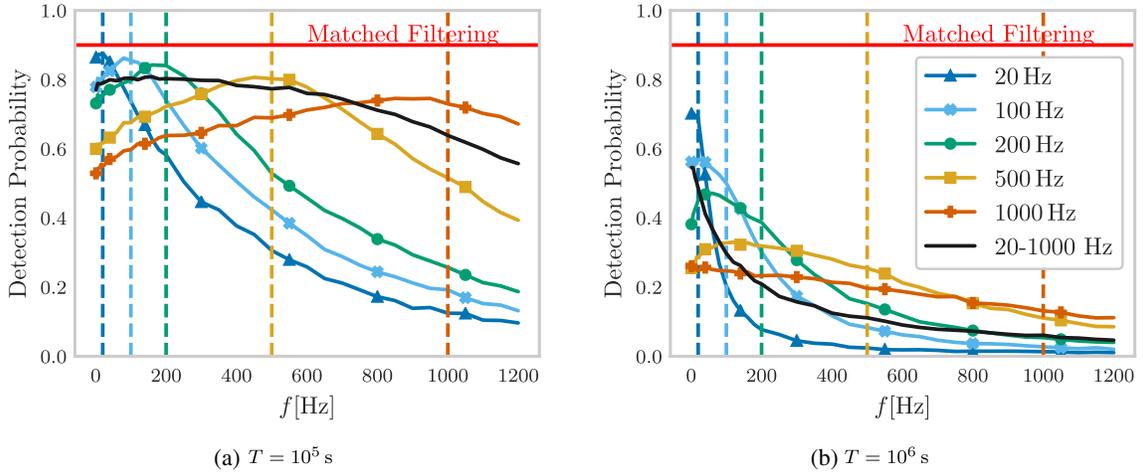

FIG. 6. Detection probability $p_{\text{det}}$ versus injection frequency $f$ for networks trained at five different frequencies and for a network trained with signals drawn from the full frequency range (solid black line). The dashed vertical lines mark the respective training frequencies for the five "specialized" networks. The horizontal dashed line represents the coherent matched-filtering detection performance.

### C. Generalization in spindown $\dot{f}$

A further interesting aspect to consider is how far in spindown $\dot{f}$ the performance network extends beyond the range that it was trained on, i.e., $\dot{f} \in [-10^{-10}, 0]$ Hz/s as given in Table I. This is shown in Fig. 7. We see that the DNN detection probability remains high even for spindowns that are 1–2 orders of magnitude larger than the training range. In particular, networks trained at higher frequencies seem to have a wider generalization range in spindown, which makes sense as they would have learned to recognize signal shapes with larger Doppler broadening, a qualitatively similar effect to having more spindown.

### D. Generalization in signal strength

Another important issue is how well the DNN generalizes for signals of different strength $\mathcal{D}$, given that we only trained each network at one specific depth $\mathcal{D}_{\text{training}}^{90\%}$, an estimate of the matched-filtering depth. The results of this test are shown in the efficiency plots of Fig. 8. We see that generally the dependence of $p_{\text{det}}(\mathcal{D})$ for the DNNs seems to be quite similar to that of matched filtering, but shifted to its overall (lower) performance level.

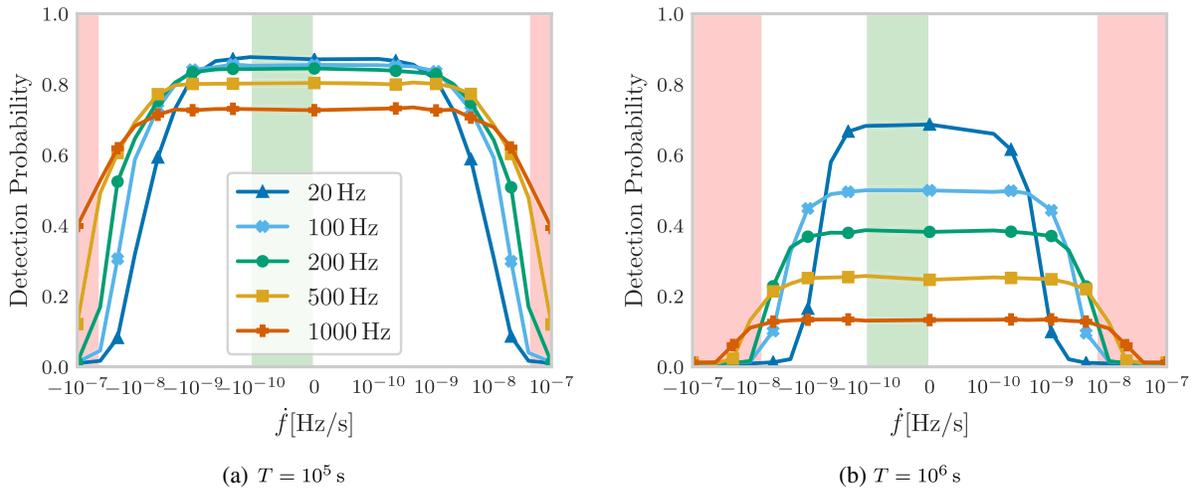

FIG. 7. Detection probability $p_{\text{det}}$ versus injected spindown $\dot{f}$ for networks trained at five different frequencies. The green shade in the middle marks the $10^{-10}$ Hz/s wide spindown band the networks were trained on. The $x$ axis is plotted as a symmetric logarithm, i.e., logarithmical for the larger negative values, linear for $|\dot{f}| < -10^{-10}$ Hz/s and logarithmical for the larger positive values. The red shades at the edges illustrate where we start losing SNR purely by the network input window being smaller than the widest signals.

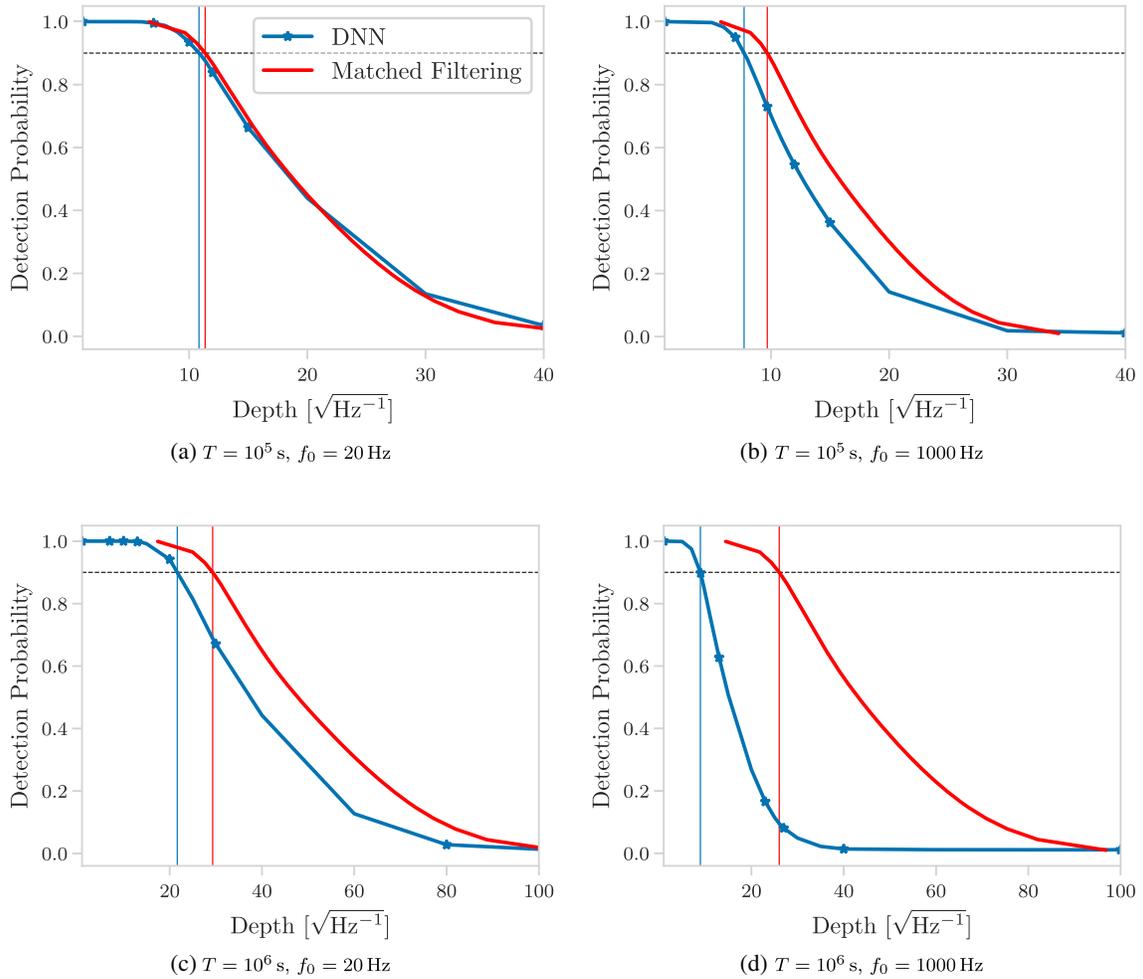

FIG. 8. Detection probability $p_{\text{det}}$ versus injection depth $\mathcal{D}$ for networks trained on the respective matched-filtering depth $\mathcal{D}^{90\%}$ (indicated by the vertical solid line with the diamond at 90%). The second vertical line with the star at 90% gives the sensitivity depth for the DNN at 90% detection probability.

TABLE VI. Sensitivity depths $\mathcal{D}_{\mathrm{DNN}}^{90\%}$ at a false-alarm level of $p_{\mathrm{fa}} = 1\%$ achieved by the network for the ten test cases. The respective matched-filter depths can be found in Table IV.

| $\mathcal{D}_{\mathrm{DNN}}^{90\%}$ [Hz$^{-1/2}$] | $f_0 = 20$ Hz | 100 Hz | 200 Hz | 500 Hz | 1000 Hz |
|---|---|---|---|---|---|
| $T = 10^5$ s | 10.8 | 10.0 | 9.5 | 8.6 | 7.7 |
| $T = 10^6$ s | 21.6 | 16.5 | 14.3 | 11.1 | 8.9 |

Conversely we also calculated the "upper limit" sensitivity depth $\mathcal{D}_{\mathrm{DNN}}^{90\%}$ where the network achieves 90% detection probability (see Table VI). These values correspond to a sensitivity loss of 5–21% (as a function of frequency) for the $T = 10^5$ s search, and 26–66% for the $T = 10^6$ s search.

### E. Timing

The total amount of computational resources needed, is another interesting point of comparison to a matched-filter search. The total search times for using the matched-filter WEAVE method on the two benchmark searches can be found in Table II.

For the DNN the total computation time consists of two parts: training time and *prediction* time (i.e., calculating one output statistic $p_{\mathrm{signal}}$ for one input data vector). The training time for the two network architectures is $\sim 1$ and $\sim 10$ d per network for the $T = 10^5$ and $T = 10^6$ s cases, respectively. Only part of this time is actually spent on training the network; another part is spent calculating the detection probability of the network every few epochs in order to monitor the progress of training.

The prediction time in comparison is almost negligible. The smaller networks for the $T = 10^5$ s cases require $\sim 3$ ms to process one input window. The larger networks for the $T = 10^6$ s cases need $\sim 10$ ms per prediction. Each search requires a different number of sliding input windows to cover the whole frequency range, and the total search time can be found in Table VII.

An important detail to note in a direct comparison between matched filtering and a pure classifier "signal" versus "noise" DNN search is that matched filtering yields far more information on outlier candidates that cross the threshold. In particular, its signal parameters will be well constrained already, allowing a follow-up search to be performed in a small region of parameter space. The DNN classifier, on the other hand, would flag input windows (of width $\Delta f_{\mathrm{IW}}$) in frequency as outliers to be followed up. Assuming we follow up two input windows per candidate, one can estimate the total expected follow-up cost (using matched filtering) as a fraction $2(\Delta f_{\mathrm{IW}}/\Delta f)p_{\mathrm{fa}}$ of the total matched-filtering cost (see Table II), where $p_{\mathrm{fa}} = 1\%$ is the false-alarm probability per $\Delta f = 50$ mHz band.

Therefore even including all the training time and assuming a matched-filter follow-up, the DNN search would still seem to require less computing power. At the present stage, however, we cannot realize this potential benefit given that our DNN search so far is far less sensitive overall.

## V. DISCUSSION

In this work we have shown that deep learning (DNNs) can in principle be used to directly search for CW signals in data, at substantially faster search times than matched filtering. For the hand-optimized network architecture studied here, the DNN detection probability (at fixed false alarm) is found to be somewhat competitive (88–73% over the full frequency range) with matched filtering (90%) for short data spans of $T \sim 1$ day, while the detection performance falls short (69–13%) for a longer data span of $T \sim 12$ days. On the plus side, the DNN search shows a surprising ability to extend further in frequency and spin-down than it was trained for, and is generally much faster in search performance than matched filtering.

In order to make this a competitive search method, we can identify a few necessary next steps:
(1) Extend to a multidetector search.
(2) Find better networks with a comparable detection probability to existing methods in Gaussian noise.
(3) Train for parameter estimation in addition to pure classification in order to reduce follow-up cost of candidates.
(4) Test how a network trained on Gaussian noise performs on real detector data. Given the network's ability to generalize, one might expect problems if non-Gaussian artifacts are identified as signals. On the other hand, training a network on real detector noise should alleviate that problem.

Overall we think that deep learning has the potential to become a useful CW search tool, but there is substantial further research and development effort required in order to achieve this.

## ACKNOWLEDGMENTS

We thank Sinéad Walsh, Maria Alessandra Papa, Marlin Schäfer and the AEI CW group for helpful comments. All WEAVE Monte Carlo simulations and all DNN training and testing were performed on the ATLAS computing cluster of the Albert-Einstein Institute in Hannover. C. M. is supported by the Science and Technology Research Council (Grant No. ST/L000946/1) and the European Cooperation in Science and Technology (COST) action CA17137.

TABLE VII. DNN computing cost (in seconds) for training, search and follow-up (using matched filtering). The respective matched-filtering cost can be found in Table II.

| Cost [s] | Training | Search | Follow-up | Total |
|---|---|---|---|---|
| $T = 10^5$ s | $4.3 \times 10^5$ | 58.8 | $2.2 \times 10^4$ | $4.5 \times 10^5$ |
| $T = 10^6$ s | $4.3 \times 10^6$ | 196 | $6.5 \times 10^7$ | $6.9 \times 10^7$ |